\documentclass[twocolumn,aps,showpacs]{revtex4}
\usepackage{graphicx}
\usepackage{fmtcount} 
\usepackage{array,multirow}
\begin{document}
\title{1D planar, cylindrical and spherical subsonic solitary waves in space electron-ion-positive dust plasma systems}
 \author{A A Mamun\footnote{Corresponding author:  mamun\_phys@juniv.edu}}
\affiliation{Department of Physics \& Wazed Mia Science Research Centre,
Jahangirnagar University, Savar, Dhaka-1342, Bangladesh}
\begin{abstract}
The space electron-ion-positive dust plasma system containing isothermal inertialess electron species, cold inertial ion species, and stationary positive (positivively charged) dust species is considered. The basic features of one dimensional (1D) planar and nonplanar subsonic solitary waves are investigated by the pseudo-potential and reductive perturbation methods, respectively. It is observed that the presence of the positive dust species reduces the phase speed of the ion-acoustic waves, and consequently supports the subsonic solitary waves with the positive wave potential in such a space dusty plasma system. It is observed that the cylindrical and spherical subsonic solitary waves significantly evolve with time, and that the time evolution of the spherical solitary waves is faster than that of the cylindrical ones. The applications of the work in many space dusty plasma systems, particularly in Earth's mesosphere, cometary tails, Jupiter's magnetosphere, etc.  are addressed.
\end{abstract}
\maketitle
The existence of the ion-acoustic  (IA) waves in a plasma medium was first predicted first by Tonks and Langmur \cite{Tonks29ia-theor} on the  basis of the fluid dynamics in 1929.  The prediction of Tonks and Langmuir \cite{Tonks29ia-theor} was then verified by Revans \cite{Revans33ia-expt} in 1933.  The well known  linear dispersion relation for the  IA waves propagating in a pure electron-ion plasma containing cold inertial ion fluid and isothermal inertia-less electron fluid  is given by 
\begin{eqnarray}
\omega=\frac{kC_i}{\sqrt{1+k^2\lambda_D^2}},
\label{IA-dispersion}
\end{eqnarray}
where $\omega=2\pi f$ and $k=2\pi/\lambda$ in which $f$ ($\lambda$) is the IA wave frequency (wavelength);  $C_i=(z_ik_BT_e/m_i)^{1/2}$ is  the IA speed in which $k_B$ is the Boltzmann constant, $T_e$ is the electron temperature, and $m_i$ is the ion mass; $\lambda_D=(k_BT_e/4\pi z_i^2n_{i0}e^2)^{1/2}$ is  the IA wave-length scale in which $n_{i0}$ ($z_i$) is the  number density (charge state) of the ion species at equilibrium, and $e$ is the magnitude of the charge of an electron.  We note that for a pure electron-ion plasma $n_{e0}=z_in_{i0}$ at equilibrium, where $n_{e0}$ is the electron number density at equilibrium.  The dispersion relation (\ref{IA-dispersion}) becomes $\omega\simeq kC_i$ for a long-wavelength limit, $\lambda\gg\lambda_D$, and  $\omega\simeq\omega_{pi}$  for a short wavelength limit, $\lambda\ll\lambda_D$,  where $\omega_{pi}=(4\pi z_i^2n_{i0}e^2/m_i)^{1/2}$ is the angular frequency of ion plasma oscillations. Thus,  the angular frequency range of the IA waves  is $0>\omega>\omega_{pi}$.

The  IA waves \cite{Tonks29ia-theor,Revans33ia-expt} are found to be modified in an 
electron-ion-negative dust plasma system theoretically 
\cite{Shukla92-dia,DAngelo93-dia,DAngelo94-dia} as well as experimentally 
\cite{Barkan96dia-expt,Merlino98dia-expt,Merlino04-dia-expt}. It has been found that the increase in number density and charge of the negative dust species enhances the phase speed of the IA waves, and consequently support the supersonic  \cite{Bharuthram92so-dia,Popel95so-dia,Nakamura01so-dia-observ,Mamun02,Mamun08,Mamun09} solitary waves (SWs). 

There are many space plasma environments, viz.
Earth's mesosphere \cite{Havnes96,Gelinas98,Mendis04}, cometary tails \cite{Horanyi96}, Jupiter's surroundings \cite{Tsintikidis96}, Jupiter's magnetosphere \cite{Horanyi93}, etc. where in addition to electron-ion plasmas, positive dust species have been observed \cite{Havnes96,Gelinas98,Mendis04,Horanyi96,Tsintikidis96,Horanyi93}.  There are three principal mechanisms by which the dust species becomes positively charged \cite{Chow93,Rosenberg95,Rosenberg96,Fortov98}. These are  photo-emission of  electrons from the dust surface induced by the flux of photons \cite{Rosenberg96},  thermionic emission of electrons from the dust grain surface by the radiative heating \cite{Rosenberg95}, and  secondary emission of electrons from the dust surface by the impact of high energetic plasma particles \cite{Chow93}.

The dispersion relation for the IA waves in an  electron-ion-positive dust plasma system (containing inertialess  isothermal electron species, inertial cold ion species,  and stationary  positive dust species) is given by
\begin{eqnarray}
\omega=\frac{1}{\sqrt{1+\mu}}\frac{kC_i}{\sqrt{1+\frac{1}{1+\mu}k^2\lambda_D^2}},
\label{MIA-dispersion2}
\end{eqnarray}
where $\mu=z_dn_{d0}/z_in_{i0}$ with $n_d$ ($z_d$) being the number density (charge state) of the positive dust species. The dispersion relation (\ref{MIA-dispersion2}) becomes $\omega\simeq kC_i/(1+\mu)$ for the long-wavelength limit (viz. $\lambda\gg\lambda_D$). The  dispersion relation 
$\omega\simeq kC_i/\sqrt{1+\mu}$ indicates that the phase speed ($\omega/k)$ decreases with the rise of the value of  $\mu$. We note that $\mu=0$ corresponds to the electron-ion plasma \cite{Tonks29ia-theor,Revans33ia-expt},  and $\mu\rightarrow\infty$ corresponds to electron-dust plasma \cite{Khrapak01,Mamun04}. Thus, $0<\mu<\infty$ is valid for the electron-ion-positive dust plasma system.

To investigate the nonlinear propagation of  the modified IA  (MIA) waves  defined by (\ref{MIA-dispersion2}),  we consider such an electron-ion-positive dust plasma system.  The nonlinear dynamics of the MIA waves  (\ref{MIA-dispersion2}) is described by 
\begin{eqnarray}
&&\frac{\partial n_i}{\partial t}
+\frac{1}{r^\nu} \frac{\partial}{\partial r} (r^\nu n_iu_i) = 0,
\label{MIA-b1}\\
&&\frac{\partial u_i}{\partial t} + u_i \frac{\partial u_i}
{\partial x} =-\frac{\partial \phi} {\partial x},
\label{MIA-b2}\\
&&\frac{1}{r^\nu}\frac{\partial}{\partial r}\left(r^\nu\frac{\partial
\phi}{\partial r}\right)  =(1+\mu)\exp(\phi)-n_i-\mu,
\label{MIA-b3}
\end{eqnarray}
where $\nu=0$ for one dimensional (1D) planar geometry and $\nu=1~(2)$ for
cylindrical (spherical) geometry; the electron species has is assumed to obey the Boltzmann law so that $n_e=\exp(\phi)$;
$n_e$ ($n_i$) is the   electron (ion) number density normalized by $n_{e0}$ ($n_{i0}$);  $u_i$ is the ion fluid speed normalized by $C_i$;  $\phi$ is the electrostatic  wave potential normalized by  $k_BT_e/e$;  $r$ and $t$ are normalized by  
$\lambda_D$ and  $\omega_{pi}^{-1}$, respectively.  The assumption of  stationary positive dust species is valid because of the mass of the positive dust species being extremely high in comparison with that of the inertial ion species.  
 
\begin{figure}[htb] 
\includegraphics[width=0.48\textwidth]{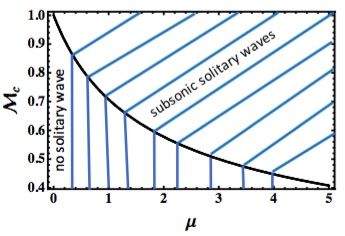}
\caption{The range of the values of $\mu$ and corresponding ${\cal M}$  for which the subsonic MIA SWs exist.} 
\label{f1}
\end{figure}

To study arbitrary amplitude MIA SWs in planar geometry ($\nu=0$ and $r=x$), we employ the pseudo-potential  approach \cite{Bernstein57PPA,Cairns95} by assuming
that all dependent variables in (\ref{MIA-b1})$-$(\ref{MIA-b3}) depend  only on a
single variable  $\xi= x - {\cal M}t$, where ${\cal M}$  is the Mach number. This transformation and steady state condition allow us to write
(\ref{MIA-b1})$-$(\ref{MIA-b3}) as
\begin{eqnarray}
&&{\cal M}\frac{d n_i}{d\xi}- \frac{d}{d\xi}(n_i u_i) =0,
\label{MIA-b4}\\
&&{\cal M}\frac{d u_i}{dl\xi}- u_i \frac{d u_i}{d\xi}=\frac{d\phi}{d\xi},
\label{MIA-b5}\\
&&\frac{d^2 \phi}{d\xi^2}=(1+\mu)\exp(\phi)-n_i-\mu
\label{MIA-b6}
\end{eqnarray}
The integration of (\ref{MIA-b4}) and (\ref{MIA-b5}) with respect to $\xi$,  and the use of appropriate boundary conditions for localized perturbations (viz. $n_i\rightarrow 1$,  $u_i\rightarrow 0$, and $\phi\rightarrow 0$ at  $\xi\rightarrow\pm\infty$) give rise to
\begin{eqnarray}
n_i=\frac{1}{\sqrt{1-\frac{2\phi}{{\cal M}^2}}}.
\label{MIA-b7}
\end{eqnarray}
Now substituting  (\ref{MIA-b7})  into (\ref{MIA-b6}), and integrating the resulting equation with respect to $\phi$, we obtain an energy integral \cite{Bernstein57PPA,Cairns95} in the form 
\begin{eqnarray}
&&\hspace*{-8mm}\frac{1}{2}\left(\frac{d\phi}{dx}\right)^2+V(\phi)=0,
\label{EI-MIA}\\
&&\hspace*{-8mm}V(\phi)= {\cal C}_0-(1+\mu)\exp(\phi)-{\cal M}^2\sqrt{1-\frac{2\phi}{{\cal M}^2}}+\mu\phi.
\label{PP}
\end{eqnarray}
The integration constant ${\cal C}_0\,[=1+\mu+{\cal M}^2]$ is chosen under the condition $V(0)=0$. 
The pseudo-potential $V(\phi)$  allow us to express as
\begin{eqnarray}
&&\left[V(\phi)\right]_{\phi=0}=0,
\label{0th-derivative}\\
&&\left[\frac{dV(\phi)}{d\phi}\right]_{\phi=0}=0,
\label{1st-derivative}\\
&&\left[\frac{d^2V(\phi)}{d\phi^2}\right]_{\phi=0}=\frac{1}{2!}\left[\frac{1}{{\cal M}^2}-(1+\mu)\right],
\label{2nd-derivative}\\
&&\left[\frac{d^3V(\phi)}{d\phi^3}\right]_{\phi=0}= \frac{1}{3!}\left[\frac{3}{{\cal M}^4}-(1+\mu)\right].
\label{3rd-derivative}
\end{eqnarray}
It is obvious from (\ref{0th-derivative})  and (\ref{1st-derivative}) that the MIA SWs exist  if and only if
$[d^2V/d\phi^2]_{\phi=0}<0$, which makes the fixed point at the origin is unstable \cite{Cairns95}, and  that 
$[d^3V/d\phi^3]_{\phi=0}\>\,(<)\,0$ for the existence \cite{Cairns95} of the MIA SWs with $\phi>0$ ($\phi<0$).  
The condition $[d^2V/d\phi^2]_{\phi=0}=0$ yields 
the critical Mach number ${\cal M}_c$ (minimum value of ${cal M}$ above which the MIA SWs exist). Thus, from  
(\ref{2nd-derivative}) we can define ${\cal M}_c$ as
\begin{eqnarray}
{\cal M}_c=\frac{1}{\sqrt{1+\mu}}.
\label{Mc}
\end{eqnarray}
The variation of ${\cal M}_c$  with $\mu$  is graphically  shown to find the range of the values of $\mu$ and corresponding ${\cal M}$  for which the subsonic MIA SWs exist. The results are displayed in  figure \ref{f1}. 
It is clear from figure \ref{f1} that  in the region above the curve (indicated by the horizontal lines)  the subsonic MIA SWs are formed, and that in the region below the curve (indicated by the vertical lines) no solitary wave exists.  On the other hand,  $[d^3V/d\phi^3]_{\phi=0} ({\cal M}={\cal M}_c)=0$  yields ${\cal S}_c$: The  subsonic MIA SWs  with
$\phi>0$ ($\phi<0$) exist  if ${\cal S}_c>0$  (${\cal S}_c<0)$, where ${\cal S}_c=\mu+2/3$.  The latter implies that  ${\cal S}_c>0$ is always valid since $0<\mu<\infty$, and that the MIA SWs  exist only with  $\phi>0$  (so, from now `SWs' will be used to mean `MIA  SWs with  $\phi>0$'). 

\begin{figure}[htb]
\includegraphics[width=0.45\textwidth]{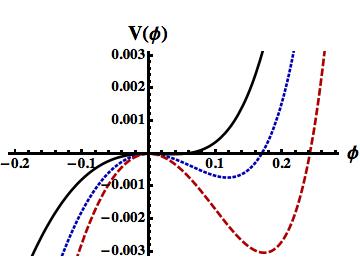} 
\caption{The potential wells corresponding to arbitrary amplitude subsonic SWs for ${\cal M}=0.9$, 
$\mu=0.1$ (solid curve),  $\mu=0.4$ (dotted curve),  and  $\mu=0.7$  (dashed curve).}
\label{f2}
\end{figure}

\begin{figure}[htb]
\includegraphics[width=0.48\textwidth]{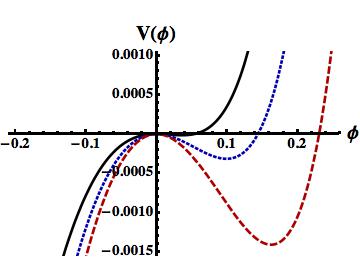} 
\caption{The potential wells corresponding to arbitrary amplitude subsonic SWs for $\mu=0.3$,  
${\cal M}=0.90$  (solid curve),  ${\cal M}=0.94$ (dotted curve),  and  ${\cal M}=0.98$  (dashed curve).}
\label{f3}
\end{figure}

We now plot $V(\phi)$ vs. $\phi$ curves to study the formation of arbitrary amplitude subsonic SWs for which the potential wells are formed in $+\phi$-axis.  The numerical results are shown in figures \ref{f2} and \ref{f3}.
The potential wells  in $+\phi$ axis in figure \ref{f2} or \ref{f3} represent the amplitude $\phi_m$ (value of $\phi$ at the point where the $V(\phi)$ vis. $\phi$ curve  crosses $+\phi$-axis) and the width ${\cal W}$ (defined as ${\cal W}=|\phi_m|/\sqrt{|V_m|}$, where $|V_m|$ is the  maximum value of $V(\phi)$ in the potential wells) of arbitrary amplitude SWs.  Thus  figures \ref{f2} and \ref{f3} indicate that the  amplitude (width) of the subsonic  SWs  increases (decrease) with the rise of the value of  $\mu$ and ${\cal M}$. 

\begin{figure}[htb]
\includegraphics[width=0.45\textwidth]{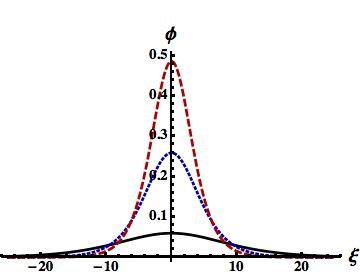} 
\caption{The small amplitude subsonic MIA SWs for ${\cal M}=0.9$,  $\mu=0.3$ (solid curve),  $\mu=0.4$ (dotted curve),  and  $\mu=0.7$  (dashed curve).}
\label{f4}
\end{figure}

\begin{figure}[htb]
\includegraphics[width=0.48\textwidth]{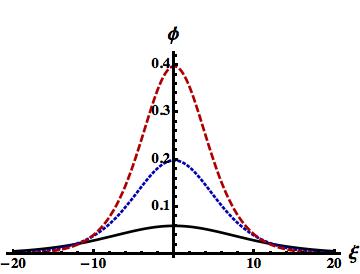} 
\caption{The small amplitude subsonic SWs for $\mu=0.5$,  ${\cal M}=0.90$  (solid curve),  
${\cal M}=0.94$ (dotted curve),  and  ${\cal M}=0.98$  (dashed curve).}
\label{f5}
\end{figure}

We now study the basic features of small amplitude subsonic SWs for which the pseudo-potential $V(\phi)$ [defined by (\ref{PP})] can be expanded as
\begin{eqnarray}
&&V(\phi)=C_2\phi^2+C_3\phi^3+\cdot\cdot\cdot, 
\label{PPEXP}\\
&&C_2=\frac{1}{2!}\left[\frac{1}{{\cal M}^2}-(1+\mu)\right],
\label{C2}\\
&&C_3= \frac{1}{3!}\left[\frac{3}{{\cal M}^4}-(1+\mu)\right].
\label{C3}
\end{eqnarray}
It is clear from (\ref{PPEXP}) that the constant and the coefficient of $\phi$ in the expansion of $V(\phi)$ vanish because of the choice of the  integration constant, and  the equilibrium charge neutrality condition, respectively.   The approximation
$V(\phi)=C_2\phi ^2+C_3\phi^3$,  which  is valid as long as $C_n\phi^n$  (where $n=4,~5,~6,\cdot\cdot\cdot$)  are negligible compared to $C_3\phi^3$), and the condition $V(\phi_m)=0$ reduce the solitary wave solution of (\ref{EI-MIA}) to
\begin{eqnarray}
\phi=\left(-\frac{C_2}{C_3}\right){\rm sech^2}\left(\sqrt{-\frac{C_2}{2}}\xi\right).
\label{sas1}
\end{eqnarray}
We have graphically represented  (\ref{sas1}) for $1>{\cal M}>{\cal M}_c$ and  $\mu>0$  to observe  the basic features of small amplitude  subsonic SWs for different values of $\mu$ and ${\cal M}$. The results are displayed in figures \ref{f4} and \ref{f5}. 
\begin{figure}[htb]
\includegraphics[width=0.48\textwidth]{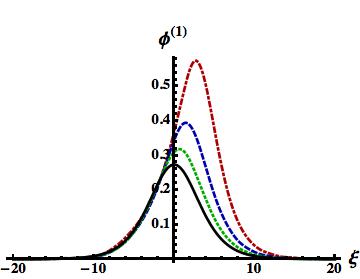} 
\caption{The time evolution of the subsonic SWs  in cylindrical ($\nu=1$) geometry for ${\cal U}_0=0.1$, $\mu=0.1$, $\tau=-20$ (solid curve),  $\tau=-15$ (dotted curve), $\tau=-10$ (dashed curve), and $\tau=-5$ (dot-dashed curve).}
\label{f6}
\end{figure}
\begin{figure}[htb]
\includegraphics[width=0.48\textwidth]{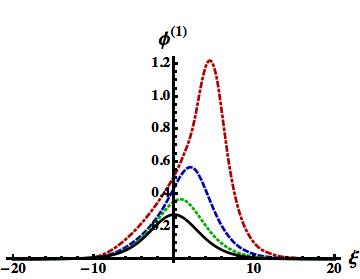} 
\caption{The time evolution of the subsonic SWs  in spherical ($\nu=2$) geometry for ${\cal U}_0=0.1$, $\mu=0.1$, $\tau=-20$ (solid curve),  $\tau=-15$ (dotted curve), $\tau=-10$ (dashed curve), and $\tau=-5$ (dot-dashed curve).}
\label{f7}
\end{figure}
It is obvious from figures \ref{f4} and \ref{f5} that stationary positive dust species supports the small amplitude subsonic SWs in an electron-ion-positive dust plasma system.  The variation of their amplitudes and width with $\mu$ and ${\cal M}$ are obvious from figures \ref{f4} and \ref{f5}.  The latter indicate that the amplitude (width) of these subsonic SWs increase (decrease) with the rise of  the  value of $\mu$ and ${\cal M}$ as we concluded from the direct analysis of the pseudo-potential  $V(\phi)$ in figures \ref{f2} and \ref{f3}.  

We finally examine the effect of nonplanar geometry on time dependent supersonic SWs 
by using the reductive perturbation method \cite{Washimi66} which requires the 
stretching the independent variables \cite{Maxon74,Mamun01}:
\begin{eqnarray}
&&\zeta=-\epsilon^{\frac{1}{2}}(r+{\cal V}_pt),
\label{str1}\\
&&\tau=\epsilon^{\frac{3}{2}}t,
\label{str2}
\end{eqnarray}
expanding the dependent variables \cite{Washimi66,Maxon74,Mamun01}:
\begin{eqnarray}
&&n_i=1+\epsilon n_i^{(1)}+\epsilon^2 n_i^{(2)}+\cdot \cdot \cdot,\\
&&u_i=0+\epsilon u_i^{(1)}+\epsilon^2 u_i^{(2)}+\cdot \cdot \cdot,\\
&&\phi=0+\epsilon \phi^{(1)}+\epsilon^2 \phi^{(2)}+\cdot \cdot \cdot,
\label{np-expan}
\end{eqnarray}
and developing equations in various powers of $\epsilon$. 
To the lowest order in $\epsilon$, (\ref{MIA-b1})$-$(\ref{MIA-b3}) yield 
\begin{eqnarray}
n_i^{(1)}= -\frac{u_i^{(1)}}{{\cal V}_p},
\label{np1}\\
u_i^{(1)}=-\frac{\phi^{(1)}}{{\cal V}_p},
 \label{np2}\\
{\cal V}_p=\frac{1}{\sqrt{1+\mu}}.
 \label{np3}
\end{eqnarray}
We note that ${\cal V}_p={\cal M}_c$, and thus the variation of ${\cal V}_p$ is shown in figure \ref{f1}. 
To the next higher order in $\epsilon$, we obtain a set of equations:  
\begin{eqnarray}
&&\hspace*{-8mm}\frac{\partial n_i^{(1)}}{\partial \tau}
-\frac{\partial}{\partial\zeta}\left[{\cal V}_pn_i^{(2)}+u_i^{(2)}+n_i^{(1)}u_i^{(1)}\right]-\frac{\nu u_i^{(1)}}{{\cal V}_p\tau}=0,
\label{np4}\\ 
&&\hspace*{-8mm}\frac{\partial u_i^{(1)}}{\partial \tau}-{\cal V}_p\frac{\partial u_i^{(2)}}{\partial \zeta}-u_i^{(1)}\frac{\partial u_i^{(1)}}{\partial \zeta}
=\frac{\partial \phi^{(2)}}{\partial \zeta},
\label{np5}\\ 
&&\hspace*{-8mm}\frac{\partial^2 \phi^{(1)}}{\partial \zeta^2}=(1+\mu)\phi^{(2)}-n_i^{(2)} +\frac{1}{2}(1+\mu)[\phi^{(1)}]^2.
\label{np6}
\end{eqnarray}
The use of (\ref{np1})$-$(\ref{np6}) gives rise to a modified Korteweg-de Vries (K-dV)
equation in the form
\begin{eqnarray}
\frac{\partial\phi^{(1)}}{\partial
\tau}+\frac{\nu}{2\tau}\phi^{(1)} + {\cal A} \phi^{(1)} \frac{\partial
\phi^{(1)}}{\partial \zeta} + {\cal B} \frac{\partial^3 \phi^{(1)}}{\partial
\zeta^3} = 0, 
\label{mK-dV}
\end{eqnarray}
where ${\cal A}$  and ${\cal B}$ are nonlinear and dispersion coefficients, and are, respectively,  given by 
\begin{eqnarray}
&&{\cal A}=\frac{3}{2\sqrt{1+\mu}}\left(\mu+\frac{2}{3}\right),
\label{np7}\\ 
&&{\cal B}=\frac{1}{2(1+\mu)^{\frac{3}{2}}}.
\label{np8}
\end{eqnarray}
We note that $(\nu/2\tau)\phi^{(1)}$ in (\ref{mK-dV}) is due to the effect of the nonplanar geometry, and that $\nu=0$ or $|\tau|\rightarrow \infty$ corresponds to a 1D
planar geometry.  Thus, for a large value of $\tau$,  and for a
frame moving with a speed ${\cal U}_0$,  the stationary solitary wave
solution of (\ref{mK-dV}) is \cite{Mamun02} 
\begin{eqnarray}
&&\phi^{(1)}=\left(\frac{3{\cal U}_0}{{\cal A}}\right){\rm
sech^2}\left[\sqrt{\frac{u_0}{4B}}(\zeta-{\cal U}_0\tau)\right].
\label{np-sol}
\end{eqnarray}
We note that $\tau<0$ means the time evolution of the SWs from past to present. But we cannot use $\tau\rightarrow 0$ because of the term  $(\nu/2\tau)\phi^{(1)}$ in (\ref{mK-dV}). 

We now numerically solve (\ref{mK-dV}) to observe how the SWs evolve with time from past to present by using (\ref{np-sol}) as the initial solitary pulse. The results for cylindrical ($\nu=1$) and spherical ($\nu=2$) geometries are shown in figures \ref{f6} and \ref{f7}, respectively.  We note that for a large value of $\tau$  (viz. $\tau=20$) we got solid curves in  figures \ref{f6} and \ref{f7}, which represent subsonic SWs in 1D planar, cylindrical, and spherical  geometries.

To summarize, we have considered  a space dusty plasma system containing  electron,  ion and positive dust species, and have identified the basic features of the subsonic SWs in such space dusty plasma systems. The results obtained from this theoretical investigation are: 
\begin{enumerate}
\item{The condition $0<\mu<\infty$ causes to form the subsonic SWs with $\phi>0$ in the space  plasma systems. This is due to the fact that the phase speed of the MIA waves in such a space  plasma systems decreases with the rise of the value of $\mu$.}  

\item{The amplitude (width) of the subsonic SWs  increases (decreases) with the rise of the value of $\mu$  because of the decrease in the Mach number with the increase in the value of $\mu$.}

\item{The amplitude of the subsonic SWs is small  because of  their existence for low Mach number (${\cal M}<1$).} 

\item{The cylindrical and spherical subsonic solitary waves significantly evolve with time, and  that the time evolution of the spherical solitary waves is faster than that of the cylindrical ones. This is due to the  time and geometry dependent extra term $(\nu/2\tau)\phi^{(1)}$ in the modified K-dV equation.}
\end{enumerate}

The disadvantage of the pseudo-potential method is that it does not allow us to observe the time evolution of the arbitrary amplitude SWs.  On the other hand, the reductive perturbation method allows as to observe the time evolution of the small amplitude SWs, but it is not valid for the arbitrary amplitude SWs.  

To overcome this limitation, one has to develop a correct numerical code, and  to solve the basic equations (\ref{MIA-b1})$-$(\ref{MIA-b3}) numerically by this numerical code. This type of numerical analysis  will be able to show the time evolution of arbitrary amplitude SWs.  This is, of  course, a challenging research problem, but beyond the scope of our present work.

We finally hope that the results of this  work should be  useful for understanding the physics of  nonlinear phenomena like subsonic solitary like structures in many space  plasma environments, viz. Earth's mesosphere \cite{Havnes96,Gelinas98,Mendis04}, cometary tails \cite{Horanyi96}, Jupiter's surroundings \cite{Tsintikidis96}, Jupiter's magnetosphere \cite{Horanyi93}, etc.

\end{document}